%Paper: hep-lat/9403003
%From: Shin-ichi Tominaga <f77497a@kyu-cc.cc.kyushu-u.ac.jp>
%Date: Fri, 4 Mar 94 18:23:55 JST
%Date (revised): Tue, 8 Mar 94 22:57:58 JST

%%%%%%%%%%%%%%%%%%%%%%%%%%%%%%%%%%%%%%%%%%%%%%%%%%%%%%%%%%%%%%%%%%%%%
% The uuencoded tar file is attached at the bottom of this text.
% In UNIX, try
%  $ uudecode { stripped file }
%  $ tar xvf all.tar
% then you will see the PostScript figure files such as fig.1 etc.
%%%%%%%%%%%%%%%%%%%%%%%%%%%%%%%%%%%%%%%%%%%%%%%%%%%%%%%%%%%%%%%%%%%%%

\input harvmac
%\draftmode
\input epsf
\Title{\vbox{\baselineskip12pt\hbox{KEK-TH-390}\hbox{KEK
Preprint 93-213}\hbox{KYUSHU-HET-17}}}
{\vbox{\centerline{Phase structure of the Higgs-Yukawa systems}
\vskip3pt
\centerline{with chirally invariant lattice fermion actions}}}

\centerline{Shin-ichi Tominaga}
\vskip0.2cm\centerline{Department of Physics, Kyushu University}
\centerline{Fukuoka 812, Japan}
\vskip0.2cm\centerline{and}
\vskip0.2cm\centerline{
Sergei V. Zenkin \footnote{$^*$} {Permanent address: Institute for Nuclear
Research of Academy of Sciences of  Russia, RU-117312 Moscow, Russian
Federation. E-mail address: zenkin@inr.msk.su}}
\vskip0.2cm\centerline{National Laboratory for High Energy Physics (KEK)}
\centerline{Tsukuba 305, Japan}

\vskip 1cm

We develop analytical technique for examining phase structure of $Z_2$,
$U(1)$, and $SU(2)$ lattice Higgs-Yukawa systems with radially frozen Higgs
fields and chirally invariant lattice fermion actions. The method is based on
variational mean field approximation. We analyse phase diagrams of such
systems with different forms of lattice fermion actions and demonstrate that
it crucially depends both on the symmetry group and on the form of the
action.
We discuss location in the diagrams of possible non-trivial fixed points
relevant to continuum physics, and argue that the candidates can exist only
in
$Z_2$ system with SLAC action and $U(1)$ systems with naive and SLAC actions.

\Date{February 28, 1994}

\newsec{Introduction}

Since discovering the first evidence of complexity of phase structure of
Higgs-Yukawa systems \ref\HN{A. Hasenfratz and T. Neuhaus, Phys. Lett. B220,
435 (1989)}, \ref\HLN{A. Hasenfratz, W. Liu, and T. Neuhaus, Phys. Lett.
B236,
339 (1990)}, \ref\LSS{I-H. Lee, J. Shigemitsu, and R. E. Shrock, Nucl.
Phys. B334, 265 (1990)} much has been done to understand them better (for a
review and list of references see \ref\Sh{J. Shigemitsu, Nucl. Phys. B
(Proc. Suppl.) 20, 515 (1991); I. Montvay, Nucl. Phys. B (Proc. Suppl.)
26, 57 (1992)},
\ref\Bea{W. Bock, A. K. De, C. Frick, J. Jers\'{a}k, and T. Trappenberg, HLRZ
J\"{u}lich 91-82, Amsterdam ITFA 91-33 (1991)}). There were studied phase
structure and continuum limits of such systems with different lattice
formulations and symmetry groups; proposed  hypotheses on possible
non-trivial
fixed points and checked some of them. There were used numerical, analytical,
and combined methods. Much has been understood about this, much is still to
be
answered. In particular, such questions as: which features of the phase
diagrams are inherent properties of  the Higgs-Yukawa model, and which are
lattice artifacts; whether non-trivial fixed points exist at intermediate
values of the Yukawa coupling; if so, what does distinguish such points on
the
phase diagrams and whether this depends on group of symmetry of the system,
have not yet definite answers.

This paper, not claiming to solve definitely those unanswered problems,
rather
pursues the following aims: to give a simple analytical technique for
examining
on the same basis the phase structure of a wide class of lattice formulations
of the Higgs-Yukawa systems with various symmetry groups; to compare the
phase
diagrams for some of such formulations for $Z_2$, $U(1)$, and $SU(2)$
symmetrical systems; to give additional arguments {\it pro et contra} those
previous results which seem to be not well-established, and to draw one's
attention to interesting points which were left being not understood well.

We shall use variational mean field approximation and analyse the systems
with
radially frozen Higgs
fields and chirally invariant lattice fermion actions possessing real
fermionic
determinants. These conditions enables us to calculate contributions of
fermionic determinant into mean field free energy --- the main problem of the
method, in a ladder approximation \ref\Ze{S. V. Zenkin, Mod. Phys. Lett. A,
to be published}, in other words, to sum up the contributions of leading and
some of those of next-to-leading orders in inverse space-time dimension
$1/D$ knowing only form of the free fermion propagator in the momentum space.
Compared with other mean field calculations \ref\Boea{W. Bock, A. K. De, K.
Jansen, J. Jers\'{a}k, T. Neuhaus, and J. Smit,
Nucl. Phys. B344, 207 (1990)}, \ref\ST{M. A. Stephanov and M. M. Tsypin,
Phys. Lett. B236, 334 (1990); Sov. Phys. JETP 70, 228 (1990)}, \ref\STs{M.
A. Stephanov and M. M.  Tsypin, Phys. Lett. B242, 432 (1991)}, \ref\STsy{M.
A.
Stephanov and M. M. Tsypin,
Phys. Lett. B261, 109 (1991)},
\ref\EK{T. Ebihara and K. Kondo, Nucl. Phys. 20 (Proc. Suppl.) 26, 519
(1992); Progr. Theor. Phys. 87, 1019 (1992)} this gives us two
advantages: to analyse the phase diagrams of the systems with a wide class of
lattice fermion actions, and, in fact, for any values
of the Yukawa coupling, including the most interesting region of intermediate
ones. On the other hand the above conditions do not allow us to discuss
interesting features revealed in a model with non-frozen Higgs field
\ref\HJS{A.
Hasenfratz, K. Jansen, and Y. Shen, Nucl. Phys. B394, 527 (1993)} and to
compare our results with those obtained with staggered fermions \LSS,
\ref\Sec{I-H.  Lee, J. Shigemitsu, and R. E. Shrock, Nucl. Phys. B330, 225
(1990); S. Aoki, I-H. Lee, D. Mustaki, J. Shigemitsu, and R. E. Shrock,
Phys. Lett. B244, 301 (1990); A. Abada and R. Shrock, Phys. Rev. D43, R304
(1991)} (except formulations of \STs, \STsy, \EK\ with a number of
staggered fermions is a multiple of
$2^{D/2}$), and with results for the Wilson fermions \Boea, \ref\GP{M. F. L.
Golterman and D. N. Petcher, Phys. Lett. B247, 370 (1990); Nucl. Phys. B359,
91
(1991)}.

By virtue of the Nielsen-Ninomiya theorem \ref\NN{H. B. Nielsen and M.
Ninomiya,
Nucl. Phys. B185, 20 (1981); Nucl. Phys. B193, 173 (1981)}, chirally
invariant
lattice fermion actions (providing they are bilinear in fermion fields and
lattice translation invariant, the conditions we shall respect) must be
either
non-Hermitean, or non-local, otherwise they involve an equal number of the
left-handed and right-handed Weyl fermions. Using our technique we cannot
take
non-Hermitean actions, being able to consider naive, non-local, and mirror
fermion actions. In all the cases the fermions coupled to the Higgs fields in
the same local way.

The outline and the results of the paper are as follows.

The systems under consideration are defined in Sect. 2. We use their usual
lattice parameterization in terms of scalar hopping parameter
$\kappa$ and Yukawa coupling $y$. In Sect. 3 we describe the method and
approximations, and obtain the explicit formulae for second order phase
transition lines $\kappa_{cr}(y)$ for the symmetry groups $Z_2$, $U(1)$, and
$SU(2)$.

In Sect. 4 we apply these formulae to the systems with naive, SLAC, a
maximally non-local (the Weyl), and a mirror fermion actions. We find that
the
phase diagram for all the systems show a kind of universality at $\kappa \geq
0$, while at $\kappa < 0$ they crucially depend on both fermion action and
symmetry group. In particular, ferrimagnetic phase does not appear in any
$Z_2$
systems, while appearing in $U(1)$ ones with the Weyl and mirror fermion
actions, and in all $SU(2)$ systems.

In Sect. 5 we make mean field estimations of fermionic propagators and
condensates along the ferromagnetic-paramagnetic phase transition lines. Both
show different behaviour in weak and strong coupling regimes in all the
systems. This enables us to speculate on locations of possible non-trivial
fixed points. We conclude that the most interesting ones can exist in $Z_2$
system with SLAC action and in $U(1)$ systems with naive and SLAC actions. We
also note impossibility of defining a chiral theory in paramagnetic phase of
mirror fermion model at strong coupling.

Sect. 6 is a summary and discussion.

\newsec{The system}

The system is defined on a hypercubic $D$-dimensional ($D$ is even) lattice
$\Lambda$ with sites numbered by $n = (n_1, ..., n_D)$, $-N/2+1 \leq n_{\mu}
\leq N/2$ ($N$ is even, and eventually tends to infinity) and with lattice
spacing $a = 1$; $\hat{\mu}$ is the  unit vector along a lattice link in the
positive $\mu$-direction. Dynamical variables are Dirac fermion fields
$\psi_n$, $\psibar_n$, and scalar fields $\Phi_n \in G$, where group $G =
Z_2$,
$U(1)$, or $SU(2)$, so that $\Phi^{\dagger}_n \Phi_n = 1$. We use the
following
representations for the group elements:
\eqn\Higgs{\Phi = T \phi \equiv \sum_{a = 0}^p T^a \phi^a,}
with real $\phi^a$ such that $\sum_{a = 0}^{p} (\phi^a_n)^2 = 1$. Hence we
have
\eqn\repphi{\eqalign{& p = 0, \quad T = 1, \quad Z_2, \cr
& p = 1, \quad T = (1, \, i), \quad U(1), \cr
& p = 3, \quad T = (1, \, i \tau^1, \, i\tau^2, \, i \tau^3), \quad SU(2).}}
We imply antiperiodic boundary conditions for the fermion and periodic for
the
scalar fields.

The system is defined by functional integrals
\eqn\sys{Z[J] = \int \prod_{n} d \Phi_n d\psi_n
d\psibar_n  \e{ -A[\Phi, \psi, \psibar] + \sum_{n, a} J^{a}_n \phi^{a}_n}}
with action
\eqn\act{A[\Phi, \psi, \psibar] = - 2\kappa \sum_{n, \mu, a} \phi^{a}_n
\phi^{a}_{n+\hat{\mu}} + \sum_{m, n} \psibar_m [\dsl_{mn} +  y (P_L
\Phi^{\dagger}_m + P_R \Phi_m) \delta_{mn}]\psi_n,}
where $d \Phi_n$ is the Haar measure on $G$; $\kappa \in (-\infty, \infty)$
is
hopping parameter; $y$ is the Yukawa coupling which without loss of
generality will be considered non-negative; $P_{L,R} = (1 \pm
\gamma_{D+1})/2$
are chiral projecting operators; $\dsl$ is a lattice Dirac operator
determining
the form of the fermion action (the systems with mirror fermions which we
consider in Sect. 4.4 is reduced to this form).

We shall consider actions with operators $\dsl$ satisfying the properties
\eqn\pr{\eqalign{ & \dsl_{mn} = -\dsl_{nm},\cr
& \dsl_{mn} = \int_{p} \e{i p (m - n)}
\sum_{\mu}  i \, \gamma_{\mu} \, L_{\mu}(p), \cr
& L^*_{\mu}(p) = L_{\mu}(p), \;\;\; L_{\mu}(-p) = - L_{\mu}(p),\cr}}
where $\int_p \equiv \int d^D p / (2 \pi)^D$, $p_{\mu} \in (-\pi, \pi)$, and
we use the Hermitean $\gamma$-matrices: $[\gamma_{\mu}, \gamma_{\nu}]_{+} = 2
\delta_{\mu \nu}$.

Action \act\ is invariant under $G \times G$ global chiral transformations
\eqn\trans{\eqalign{ &\psi_n \rightarrow (h_L P_L + h_R P_R)
\psi_n,\cr
&\psibar_n \rightarrow \psibar_n (P_R h^{\dagger}_L+ P_L h^{\dagger}_R), \cr
&\Phi_n \rightarrow h_L \, \Phi_n \, h^{\dagger}_R,}}
with $h_{L, R} \in G$.

\newsec{Method and approximations}

To analyze the phase diagrams of the system we use the variational mean field
approximation \ref\DZ{J.-M. Drouffe and J.-B. Zuber, Phys. Rep. 102, 1
(1983)}
which becomes applicable to
\sys\ after integrating out the fermions
\eqn\Z{\eqalign{ Z[J] & = \e{- W[J]} \cr
& = \int \prod_{n} d \Phi_n
\e{ 2 \kappa \sum_{n, \mu, a} \phi^{a}_n  \phi^a_{n+\hat{\mu}} + \ln \det
\,[\dsl +  y \tilde{\Phi}] + \sum_{n, a} J^{a}_n \phi^{a}_n},}}
where $\tilde{\Phi} \equiv (P_L \Phi^{\dagger} + P_R \Phi)$.
Then for free energy of the system $F = W[0]$ the method yields inequality
\eqn\MF{\eqalign{F \leq F_{MF} = \inf_{h^{a}_n} \biggl[& \sum_n u(h_n) +
\sum_{n, a} h^{a}_n \langle \phi^{a}_n \rangle_{h} \cr
& - \vev{ 2 \kappa \sum_{n, \mu, a} \phi^{a}_n \phi^{a}_{n+\hat{\mu}} +
\ln \det \,[\dsl +  y \tilde{\Phi}]}_h \biggr],}}
where $h^{a}_n$ is the mean field with radial component $h_n = \bigl[\sum_a
(h^{a}_n)^2 \bigr]^{1/2}$,
\eqn\u{\eqalign{u(h_n) & = - \ln \int d\Phi_n \e{\sum_{a} h^{a}_n \phi^{a}_n}
\cr & = - \ln \cosh h_n = -{1 \over 2} h^{2}_n + {1 \over 12} h^{4}_n
+ O(h^{6}_n), \quad Z_2,
\cr & = -\ln I_0(h_n) = -{1 \over 4} h^{2}_n + {1 \over 64} h^{4}_n
+ O(h^{6}_n), \quad U(1),
\cr & = -\ln {2 \over h_n} I_1(h_n) = -{1 \over 8} h^{2}_n + {1 \over 384}
h^{4}_n + O(h^{6}_n), \quad SU(2),}}
is the test free energy per lattice site [$I_{\nu}(z)$ is the modified Bessel
functions], and
\eqn\vevh{\vev{O[\phi]}_h = \e{\sum_n u(h_n)}
\int \prod_n d\Phi_n \; O[\phi] \; \e{\sum_{n, a} h^{a}_n \phi^{a}_n}.}

The inequality \MF\ is expected to tend to equality in the limit of $D
\rightarrow \infty$ \DZ. So, we can get some idea of the system at $D = 4$
studying $F_{MF}$. From \u\ it immediately follows that
\eqn\cor{\eqalign{\vev{\phi^{a}_n}_h & = -{\partial \over \partial
h^{a}_n} u(h_n) \cr
& = h_n + O(h^{3}_n), \quad Z_2, \cr
& = {1 \over 2} h^{a}_n + O(h^{3}_n), \quad U(1), \cr
& = {1 \over 4} h^{a}_n + O(h^{3}_n), \quad SU(2); \cr}}

\eqn\corre{\eqalign{\vev{\phi^{a}_m\,\phi^{b}_n}_h& = {\partial \over
\partial h^{a}_m} u(h_m) \;{\partial \over \partial h^{b}_n} u(h_n) +
\delta _{m n} \; {\partial
\over
\partial h^{a}_n} {\partial \over \partial h^{b}_n} u(h_n) \cr
& = h_m h_n + \delta _{m n} (1 - h^{2}_n) + O(h^{4}_n), \quad Z_2, \cr
& = {1 \over 4} h^{a}_m h^{b}_n + {1 \over 2} \delta_{m n} \bigl[
\delta^{a b} (1 - {1 \over 8} h^{2}_n) -{1 \over 4} h^{a}_n
h^{b}_n\bigr] + O(h^{4}_n), \quad U(1),
\cr & = {1 \over 16} h^{a}_m h^{b}_n + {1 \over 4} \delta_{m n} \bigl[
\delta^{a b} (1 - {1 \over 24} h^{2}_n) -{1 \over 12} h^{a}_n
h^{b}_n\bigr] + O(h^{4}_n), \quad SU(2); \cr}}
and so on. Therefore the main problem is calculation of expectation
value $\vev{\ln \det \,[\dsl +  y \tilde{\Phi}]}_h$. In our case it has the
following representations
\eqn\d{\eqalign{\vev{\ln &\det \,[\dsl +  y \tilde{\Phi}]}_h \cr
& = \ln \det [\dsl] - {1 \over 2}
\sum^{\infty}_{n = 1} {1 \over n} \, y^{2n} \sum_{i_1, \ldots ,i_{2n}} \tr\,
\bigl[ (\dsl^{-1}_{i_1\,i_2} \,\dsl^{-1}_{i_2\,i_3}
\cdots \dsl^{-1}_{i_{2n}\,i_1})
\vev{\Phi^{\dagger}_{i_1} \Phi_{i_2} \cdots \Phi_{i_{2n}}}_h \bigr]\cr
& = c_g \, 2^{D/2} N^D \ln y \cr
& \quad \quad \quad \quad - {1 \over 2}
\sum^{\infty}_{n = 1} {1 \over n} \, {1 \over y^{2n}} \sum_{i_1,
\ldots ,i_{2n}}
\tr\,
\bigl[ (\dsl_{i_1\,i_2} \,\dsl_{i_2\,i_3} \cdots \dsl_{i_{2n}\,i_1})
\vev{\Phi^{\dagger}_{i_1} \Phi_{i_2} \cdots \Phi_{i_{2n}}}_h \bigr], \cr
}}
where $\tr$ stands for the trace over spinorial indices, as well as for group
ones for $SU(2)$ in which case factor $c_g = 2$, otherwise $c_g = 1$, and it
has been taken into account that the trace of an odd number of
$\gamma$-matrices vanishes; correlators of the Higgs fields in \d\ in terms
of their real components read as
\eqn\vp{\vev{\Phi^{\dagger}_{i_1} \Phi_{i_2} \cdots \Phi_{i_{2n}}}_h =
\sum_{a_1, \ldots, a_{2n}} {T^{a_1}}^{\dagger} T^{a_2} \cdots T^{a_{2n}}
\vev{
\phi^{a_1}_{i_1} \phi^{a_2}_{i_2} \cdots \phi^{a_{2n}}_{i_{2n}}}_h.}

We consider $F_{MF}$ on translation invariant ansatz for $h^{a}_n$
\eqn\h{ h^{a}_n = h^{a} + \epsilon_n
h^{a}_{st}, \quad
\epsilon_n = (-1)^{\sum_{\mu} n_{\mu}}.}
Then the mean field equations are reduced to
\eqn\mfe{{\del \over \del h} F_{MF} = 0, \quad {\del \over \del h_{st}}
F_{MF}
= 0.}
In view of \cor, the solutions of Eqs. \mfe, $h^*$ (``magnetization") and
$h_{st}^*$ (``staggered magnetization"), in fact are the order parameters
distinguishing the ferromagnetic (FM):
$h^{*} \neq 0$, $h_{st}^{*} = 0$;  antiferromagnetic (AM):
$h^{*} = 0$, $h_{st}^{*} \neq 0$; paramagnetic (PM): both are zero, and
ferrimagnetic (FI): both are nonzero, phases in the system.

Further simplification comes from the observation (see, for example
\ST), that as the values $h = 0$, $h_{st} = 0$ are always solutions of Eqs.
\mfe, and, therefore, the second order phase transition lines are determined
by
equation
\eqn\tl{ {\del^2 \over \del h^2} F_{MF} \, {\del^2 \over \del h_{st}^2}
F_{MF} - {\del^2 \over \del h \, \del h_{st}} F_{MF} =
0}
at $h = 0$, $h^{*}_{st}$ and $h^*$, $h_{st} = 0$, in order to find the
critical
lines not far from the PM phase, it is sufficient to know $F_{MF}$ to terms
of
order of $h_n^{2}$.

If the problem could be solved exactly both of two representations \d\ of the
fermionic determinant would yield the same answer. But correlations of
$\phi{^,}$s at coinciding arguments make the problem unsolvable exactly.
Indeed, the contributions of order of $h_n^{2}$ to \d\ come not only from
terms
$\propto u'^2$ [first terms in Eq. \corre], but also from terms of any orders
in $u''$ [terms proportional to $\delta_{m\,n}$ in \corre], as well as from
higher correlators. Some of such contributions shown schematically in Fig. 1.
Therefore, we are forced to use some approximations, and, in particular, to
use two representations of \d\ separately for ``weak" and ``strong" coupling
regimes of $y$, though the exact meaning of this can only be clear {\it a
posteriori}.

Our approximation involves summing up all diagrams of Fig.1 ($a$) (proper
ladder diagrams) and ($b$) (crossed ladder diagrams), so  we  call it
ladder approximation. Diagrams ($a$) and ($b$) correspond to
contributions to $\vev{\ln \det \,[\dsl +  y \tilde{\Phi}]}_h$ of the form
$u'_{i_1} u'_{i_{n+1}} \, (u'')^{2n-2}\,\delta_{i_2 \,i_{2n}} \cdots
\delta_{i_{n} \,i_{n+2}}$, and
$(u'')^{2n} \, \delta_{i_1\, i_{n+1}}  \delta_{i_2\, i_{2n}} \cdots
\delta_{i_{n}\, i_{n+2}}$, respectively.

Terms $\propto h \, h_{st}$ do not contribute to $F_{MF}$ (due to momenta
conservation). Hence, Eq.
\tl\ in neighbourhood of the PM phase turns to pair independent equations
\eqn\tli{{\del^2 \over \del h^2} F_{MF}\bigl|_{h = h_{st} = 0} =
0,
\quad {\del^2 \over \del h_{st}^2} F_{MF}\bigl|_{h = h_{st} = 0} = 0.}
Then, using properties \pr\ of the Dirac
operator, formulae \corre, and the properties of operators $T$:
\eqn\prt{\eqalign{\sum_{a = 0}^p \, {T^a}^{\dagger} T^b {T^a}^{\dagger}
& = 1, \quad Z_2, \cr
& = 0, \quad U(1), \cr
& = - 2 {T^b}^{\dagger}, \quad \tr \, {T^a}^{\dagger} T^b = 2 \delta^{a b},
\quad SU(2),}}
we find that the second order phase transition lines for the system \sys,
\act\ in our approximation are determined by the expressions:

$Z_2:$
\eqn\res{\eqalign{
& \kappa^{FM(W)}_{cr}(y) = {1 \over 4D} \biggl\{ 1 - 2^{D/2} \biggl[ { y^2
G^W(0)
\over 1 + y^2 G^W (0)} \cr
& \quad \quad \quad \quad \quad \quad \quad \quad \quad \quad \quad -
\int_q \biggl( { y^2 G^W (q) \over (1 + y^2 G^W (q))^2} + y^4 {G^W}^2(q)
\biggr) \biggr] \biggr\},
\cr  &\kappa^{AM(W)}_{cr}(y) = -{1 \over 4D}  \biggl\{ 1 - 2^{D/2} \biggl[ {
y^2 G^W (\pi)
\over 1 + y^2 G^W (\pi)} \cr
& \quad \quad \quad \quad \quad \quad \quad \quad \quad \quad \quad -
\int_q \biggl( { y^2 G^W (q) \over (1 + y^2 G^W (q))^2} + y^4
{G^W}^2(q) \biggr) \biggr] \biggr\},\cr
&\kappa^{FM(S)}_{cr}(y) = {1 \over 4D} \biggl\{ 1 - 2^{D/2} \biggl[ { G^S (0)
\over y^2 + G^S (0)} \cr
& \quad \quad \quad\ \quad \quad \quad \quad \quad \quad \quad \quad -
\int_q
\biggl({ y^2 G^S (q) \over (y^2 + G^S (q))^2} + {1
\over y^4} {G^S}^2(q)\biggr)\biggr]\biggr\},\cr
&\kappa^{AM(S)}_{cr}(y)
= -{1 \over 4D} \biggl\{ 1 - 2^{D/2} \biggl[ { G^S (\pi) \over y^2 + G^S
(\pi)} \cr
& \quad \quad \quad \quad \quad \quad \quad \quad \quad \quad \quad -
\int_q \biggl( { y^2 G^S (q) \over (y^2 + G^S (q))^2}+ {1 \over y^4}
{G^S}^2(q)\biggr)\biggr]\biggr\}; \cr }}

$U(1):$
\eqn\resu{\eqalign{
& \kappa^{FM(W)}_{cr}(y) = {1 \over 2D} \biggl[ 1 - 2^{D/2} \,{ y^2 \over
2}\,
G^W(0)
\biggr],
\cr
&\kappa^{AM(W)}_{cr}(y) = -{1 \over 2D} \biggl[ 1 - 2^{D/2}\, { y^2 \over
2}\,
G^W (\pi)
\biggr],\cr
&\kappa^{FM(S)}_{cr}(y) = {1 \over 2D} \biggl[ 1 - 2^{D/2}\, {1 \over 2
y^2}\,
G^S (0)], \cr
&\kappa^{AM(S)}_{cr}(y)
= -{1 \over 2D} \biggl[ 1 - 2^{D/2}\, {1 \over 2 y^2} \, G^S (\pi)\biggr];\cr
}}

$SU(2):$
\eqn\resul{\eqalign{
& \kappa^{FM(W)}_{cr}(y) = {1 \over D} \biggl\{ 1 - 2^{D/2} \biggl[ { y^2
G^W(0)
\over 2 - y^2 G^W (0)} \cr
& \quad \quad \quad \quad \quad \quad \quad \quad \quad \quad \quad - \int_q
\biggl( { 2 y^2 G^W (q) \over (2 - y^2 G^W
(q))^2} - {1
\over 4} y^4 {G^W}^2(q)\biggr)\biggr]\biggr\},
\cr  &\kappa^{AM(W)}_{cr}(y) = -{1 \over D} \biggl\{ 1 - 2^{D/2} \biggl[ {
y^2
G^W (\pi)
\over 2 - y^2 G^W (\pi)} \cr
& \quad \quad \quad \quad \quad \quad \quad \quad \quad \quad \quad - \int_q
\biggl( { 2 y^2 G^W (q) \over (2 - y^2 G^W
(q))^2}  - {1 \over 4} y^4
{G^W}^2(q)\biggr)\biggr]\biggr\},\cr
&\kappa^{FM(S)}_{cr}(y) = {1 \over D} \biggl\{ 1 - 2^{D/2} \biggl[ { G^S (0)
\over 2 y^2 - G^S (0)} \cr
& \quad \quad \quad \quad \quad \quad \quad \quad \quad \quad \quad - \int_q
\biggl({ 2 y^2 G^S (q) \over (2 y^2 - G^S (q))^2} -
{1
\over 4 y^4} {G^S}^2(q)\biggr)\biggr]\biggr\},
\cr  &\kappa^{AM(S)}_{cr}(y)
= -{1 \over D} \biggl\{ 1 - 2^{D/2} \biggl[ { G^S (\pi) \over 2 y^2 - G^S
(\pi)} \cr
& \quad \quad \quad \quad \quad \quad \quad \quad \quad \quad \quad -
\int_q \biggl( { 2 y^2 G^S (q) \over (2 y^2 - G^S (q))^2} - {1 \over 4 y^4}
{G^S}^2(q)\biggr)\biggr]\biggr\}; \cr }}
where
\eqn\g{\eqalign{&G^W (q) =  \int_p {L(p) L(p+q) \over L^2 (p) L^2(p+q)},
\cr &
G^S (q) = \int_p L(p) L(p+q), \cr
&\int_q \equiv \int {d^D q \over {(2 \pi)^D}}, \quad q_{\mu} \in (-\pi, \pi].
\cr}}

In \res\ -- \resul\ functions $\kappa^{FM}(y)$ describe the critical lines
between the FM and PM phases, and $\kappa^{AM}(y)$ those between the AM and
PM
phases, while superscripts $W$ and $S$ mean strong and weak coupling regimes,
respectively. The contributions to \res\ -- \resul\ which are proportional to
$G(0)$ or $G(\pi)$ come from the diagrams of Fig. 1($a$), while the
integral terms from those of Fig. 1($b$).  The second terms in
the integrands take into account that the first crossed ladder diagram enters
with a factor $1$, rather than $2$ (in preprint version of \Ze\ and in
\ref\Zen{S. V. Zenkin, KYUSHU-HET-8 (1993), hep-lat/9309010, unpublished}
such
terms in the equations for $Z_2$ system were erroneously missed). In the case
of $U(1)$ due to the specific property of operator $T$ \prt\ both ladder and
crossed ladder diagrams vanish, and all the contribution of the fermionic
determinant is due to the first diagram of Fig. 1($a$). In the case of $n_f$
fermions couple to the Higgs fields with the same $y$, factor $n_f 2^{D/2}$
appears instead of $2^{D/2}$ in Eqs. \res\ -- \resul.

The diagrams of Fig. 1($a$) are generalization of ``double chain"
diagrams of Refs. \ST, \STsy\ to any configurations of the same topology.
These
ladder diagrams coincide with the double chain ones only in the case of
strict
locality of the Dirac operator (or its inverse operator). In fact, this is
the
case only for the strong coupling regime for the systems with naive action
(see Sect. 4.1). Crossed ladder diagrams of Fig. 1($b$) are generalization
of the double chain diagrams with coinciding ends, which have not been taken
into account in previous calculations. Although the crossed ladder diagrams
are of $O(D^{-1})$ compared with the proper ones, and can be neglected at
``weak" as well as at ``strong" couplings, they can become dominating at
``intermediate" ones, when $y$ is close to the singular points of the
integrands in
\res\ and
\resul.

are basically determined by four constants $G^W(0)$, $G^W(\pi)$, $G^S(0)$,
and
$G^S(\pi)$, which, in their turn, are determined by the form of the free
fermion propagators. It is these constants that determine the singular
points of the expressions for $Z_2$ and $SU(2)$, thereby determining domains
of the weak  and strong coupling regimes in these cases. As $G(0) > 0$,
$G(\pi) < 0$, these domains are:

$Z_2:$
\eqn\do{ y < y^W \equiv |G^W(\pi)|^{-{1 \over 2}}, \quad y > y^S \equiv
|G^S(\pi)|^{1
\over 2};}

$SU(2):$
\eqn\dom{ y < y^W \equiv \bigl[{1 \over 2} G^W(0)\bigr]^{-{1 \over 2}}, \quad
y
> y^S \equiv
\bigl[{1 \over 2} G^S(0)\bigr]^{1 \over 2}.}
There can be the following possibilities:

($i$) $y^W < y^S$. Although we have no analytical expressions
describing the system in the region $y^W \leq y \leq y^S$, we can get, in
fact, complete picture of the phase diagrams if the integral terms diverge at
the points  $y^W$ and $y^S$ and no FI phases appear. These are the cases of
$Z_2$ systems with naive and mirror fermions [Figs 2($a$), ($d$)].

($ii$) $y^W > y^S$. The domains of the weak and strong coupling regimes are
overlapped, so we can continue the lines $\kappa^W (y)$ and $\kappa^S(y)$
until they intersect each other. In these cases we know the phase diagrams in
fact at any $y$ [an example is $U(1)$ system with naive action, Fig. 3($a$)]
(see, however, remark at the end of Sect. 6).

($iii$) Lines $\kappa^{FM}$ intersect lines $\kappa^{AM}$ forming the FI
phases
before cases ($i$) or ($ii$) are realized. In such a case formulae \res\ --
\resul\ describe the phase diagrams only in a neighbourhood of the PM phases
(examples are all $SU(2)$ systems, Figs. 4). The important fact is that the
lines together with their first derivatives are smooth in the points of
intersection. This follows from Eqs. \tli.

For $U(1)$ systems we continue lines $\kappa^W (y)$ and $\kappa^S(y)$ until
either case ($ii$) or ($iii$) is realized.  In this case, however, such a
procedure need to be justified.  The natural and simple way is
checking that the intersections of the curves occur at the points at which
quantities $(y^2/2) G^W$ and $G^S/(2 y^2)$ are obviously less than $1$.

The proper ladder diagrams give the contributions beginning from the order
$O(y^{\pm 2})$, the crossed ladder ones from
$O(y^{\pm 4})$ (note that $\int_q G(q) = 0$). Contributions of other diagrams
[Fig. 1($c$)] come into play in higher orders in $y^{\pm 2}$, at least from
the
order of $y^{\pm 6}$. So, we expect that these contributions are non-singular
and numerically suppressed.

\newsec{Phase diagrams}

In this section we apply formulae \res\ -- \resul\ to four-dimensional
systems
with different chirally invariant formulations of fermions on a lattice. All
the phase diagrams are shown for $n_f = 1$, unless other is indicated.

\subsec{Naive fermion action}

In this case operator $\dsl$ is local
\eqn\naive{\eqalign{&\dsl_{mn} = \sum_{\mu} \gamma_{\mu} \,{1\over 2}
(\delta_{m + \hat{\mu} \; n}  - \delta_{m - \hat{\mu} \; n}),\cr
&L_{\mu}(p) = \sin p_{\mu},\cr}}
and therefore produces species doubling. The system is invariant under
transformations: $(\psi, \psibar)_n$ $\rightarrow \exp(i \epsilon_n \pi /4)
(\psi, \psibar)_n$, $\phi_n \rightarrow \epsilon_n \phi_n$, $\kappa
\rightarrow
-\kappa$, $y \rightarrow -i y$ \ST, that, in particular, results in
$G(\pi) = -G(0)$. Numerically one has
\eqn\gnn{G^W(0) = - G^W(\pi) \approx 0.620, \quad G^S(0) = - G^S(\pi) = 2.}

Phase diagrams for $Z_2$, $U(1)$, and $SU(2)$ at $n_f = 2$ are shown in Figs.
2($a$), 3($a$), and 4($a$), respectively. Some comments follow.

$Z_2$: Fig. 2($a$). Domains of the weak and strong coupling regimes are
determined by
\eqn\dswnz{y^W \approx 1.27, \quad y^S \approx 1.41.}
The lines $\kappa^{FM}_{cr}(y)$ and $\kappa^{AM}_{cr}(y)$ do not intersect
each
other, so no FI phase appears, the case (i) being realized. The lines form
two
disconnected domains with PM phases, as well as with AM phases; the funnel
with FM phase is formed by logarithmic dropping of the critical lines near
the
points $y^W$ and $y^S$: $\kappa^{FM(W)}_{cr} \propto \ln[1 + y^2 G^W(\pi)]$,
$\kappa^{FM(S)}_{cr} \propto \ln[1 + G^S(\pi)/y^2]$, and therefore extends up
to $\kappa \rightarrow - \infty$.

$U(1), \, n_f = 2$: Fig. 3($a$). The lines form three connected domains with
FM,
PM, and
AM phases; no FI phase appears. These features are hold for any $n_f$. Both
curves intersect each others (point A and its counterpart on PM-AM line) at
$y
\approx 1.34$, where the construction is justified. The result agrees with
previous analytical ones
\STs, \EK, but disagrees with numerical results of Ref. \HLN, where an
evidence
of FI phase
has been revealed. This fact, as well as point A which is a candidate for
non-trivial fixed point, is discussed in Sect. 5 and 6.

$SU(2), \, n_f = 2$: Fig. 4($a$). Domains of the weak and strong coupling
regimes are overlapped:
\eqn\dswnsu{y^W \approx 1.80, \quad y^S = 1,}
the case ($iii$) being realized. The FI phase appears from $n_f = 1$
extending with $n_f$. In the domain of applicability of our formulae (solid
lines) the phase diagram is in a quantitative agreement with the Monte Carlo
results of Ref. \Boea.

\subsec{SLAC fermion action}

This action is defined by operator $\dsl$ of the form \ref\SLAC{S. D. Drell,
M.
Weinstein, S.
Yankielowicz, Phys. Rev. D14, 487 (1976)}
\eqn\SLAC{\eqalign{&\dsl_{mn} = \sum_{\mu} \gamma_{\mu} \, \sum_{l
> 0}(-1)^{l+1} {1\over l} \, (\delta_{m + l\hat{\mu} \; n}  - \delta_{m -
l\hat{\mu} \; n}),\cr   &L_{\mu}(p) = p_{\mu}, \quad p_{\mu} \in (-\pi, \pi).
\cr}}  It represents an action with a moderate non-locality as $\dsl_{mn}$
drops with distance like $|m-n|^{-1}$. In this case we have
\eqn\gnslac{G^W(0) \approx 0.109, \quad G^W(\pi) \approx - 0.0544, \quad
G^S(0)
= {4 \over 3} \pi^2, \quad G^S(\pi) = -{2 \over 3} \pi^2.}

Phase diagrams are shown in Figs. 2($b$), 3($b$), and 4($b$).

$Z_2$: Fig. 2($b$). Domains of the weak and strong coupling regimes are
overlapped:
\eqn\dswslac{y^W \approx 4.29, \quad y^S \approx 2.57.}
There are three connected domains with FM, PM, and AM phases, and no FI
phase.
These features are hold for any $n_f$.

$U(1)$: Fig. 3($b$). The phase diagram for $n_f = 1$ looks like in the
case of $Z_2$. But for $n_f \geq 2$ the FI phase appears. Points of
intersection of the critical lines located at $y \approx 3.31$, where the
construction is justified too.

$SU(2)$: Fig. 4($b$). Domains of the weak and strong coupling
regimes coincide with \dswslac\ (for $y^W$ within calculational errors).
FI phase appears for any $n_f \geq 1$.

\subsec{Weyl fermion action}

This action is defined by the finite dimensional approximation of functional
integrals for Weyl quantization \ref\Zenk{S. V. Zenkin, Mod. Phys. Lett. A6,
151
(1991)}. In this case we have
\eqn\Weyl{\eqalign{&\dsl_{mn} = \sum_{\mu} \gamma_{\mu}\, \sum_{l >
0}(-1)^{l+1}\, 2\, (\delta_{m + l\hat{\mu} \; n}  - \delta_{m - l\hat{\mu} \;
n}),\cr
&L_{\mu}(p) = 2 \tan {1\over2}p_{\mu}.\cr}}
This is a maximally non-local action in the sense that $\dsl_{mn}$ does not
drop at all with increasing $|m-n|$. But a remarkable fact is that this
action
can be transform to a local form if we introduce variables $\psi^d$ defined
on
the centres of D-cells of the lattice, i.e. on the sites of the dual lattice,
leaving fields $\psibar$ being defined on sites of the original one. Then
change of variables, which in momentum space looks like
$\psi_p = F(p) \psi^d_p$, with $F(p) = \prod_{\mu} \cos {1\over2}
p_{\mu}$, leads to a local action with
$L_{\mu}(p) = 2 \sin {1\over2} p_{\mu} \prod_{\nu \neq \mu} \cos
{1\over2} p_{\nu}$. Although now $L_{\mu}(p)$ has additional zeroes at the
Brillouin zone boundary, the system of course is not changed:
contributions of the additional species to the partition function are
canceled
by Jacobian coming from the change of variables, while their coupling to the
Higgs field is suppressed by the factor $F(p)$ (in this point this looks
similar to the Zaragoza proposal \ref\Aea{J. L. Alonso, J. L. Cort\'{e}s, F.
Lesmes, Ph. Boucaud, and E. Rivas, Nucl. Phys. B (Proc. Suppl.) 29B,C, 171
(1992)}).

The more non-locality of the action, the less $|G^W|$, and the grater
$|G^S|$.
In this extremal case we have (at $N \rightarrow \infty$)
\eqn\gnw{G^W(0) \approx 0.0450, \quad G^W(\pi) \approx - 0.00739, \quad
G^S(0)
\rightarrow \infty, \quad G^S(\pi) = -16.}

The divergence of $G^S(0)$ means that in this case terms of the strong
coupling
expansion diverge, and cannot be summed up into the finite expression in
the ladder approximation (in \Zen, because of missing the second term in the
integrand of \res, the wrong conclusion has been made on this point). Thus,
we
can analyse the phase diagram only in the weak coupling regime; the results
are in Figs. 2($c$), 3($c$), and 4($c$).

$Z_2$: Fig. 2($c$). Formally domains of the weak and strong coupling regimes
are overlapped:
\eqn\dswwz{y^W \approx 11.6, \quad y^S = 4.}
Despite a tendency, no FI phase appears.

$U(1)$: Fig. 3($c$). The FI phase appears at $y \approx 5.16$, where our
condition is satisfied.

$SU(2)$: Fig. 4($c$). In this case formally we have
\eqn\dswws{y^W \approx 6.67, \quad y^S \rightarrow \infty.}
The phase diagram is similar to that of $U(1)$ case, however, in view of
\dswws\ it looks plausible that at $y > y^W$ only the FM phase exists.

\subsec{Mirror fermion action}

We consider the simplest variant of the mirror fermion action \ref\Mm{I.
Montvay, Phys. Lett. 199B, 89 (1987); Nucl. Phys. B (Proc. Suppl.) 29B,C,
159 (1992)
} with zero (bare) mixing parameter between fermion field $\psi$ and its
mirror counterpart
$\chi$ and with only $\chi$ coupled to the Higgs field
\eqn\mf{\eqalign{A = & \sum_{m, n} \big[\psibar_m \, (\dsl^{N}_{m\,n} \,
\psi_n
+ W_{m\,n}\,{\chi}_n ) + \overline{\chi}_m \, (\dsl^{N}_{m\,n} \, \psi_n +
W_{m\,n}\,{\psi}_n )\big]  \cr
+ & \sum_{n} \,y \, \overline{\chi}_n (P_L
\Phi_m + P_R \Phi^{\dagger}_m) \chi_n, \cr}}
where $\dsl^{N}$ is Dirac operator for naive fermions \naive, while $W$ is
the Wilson operator
\eqn\wil{W_{m\,n} = -{1 \over 2} (\delta_{m + \hat{\mu} \; n}  + \delta_{m -
\hat{\mu} \; n} - 2\,\delta_{m\,n}).}
The action has the mirror symmetry
\eqn\mfs{\eqalign{ & \psi_n \rightarrow (h_L P_L + h_R P_R) \psi_n, \quad
\psibar_n
\rightarrow \psibar_n (P_R h^{\dagger}_L + P_L h^{\dagger}_R), \cr
& \chi_n \rightarrow (h_R P_L + h_L P_R) \chi_n, \quad \overline{\chi}_n
\rightarrow
\overline{\chi}_n (P_R h^{\dagger}_R + P_L h^{\dagger}_L), \cr
& \Phi_n \rightarrow h_L \Phi_n h^{\dagger}_R.}}
We can apply our formulae to this system after integrating out $\psi$. Then,
as $\psi$ does not couple to $\phi$ and therefore its determinant is an
irrelevant constant, we come to effective non-local action in terms of fields
$\chi$ and $\phi$ of the form of \sys\ -- \pr\ with

\eqn\ml{\eqalign{&\dsl_{mn} = [\dsl^N - W\,(\dsl^N)^{-1} W ]_{mn}, \cr
&L_{\mu}(p) = \sin p_{\mu} \, \biggl[ 1 + {\bigl(\sum_{\nu} (
1 - \cos p_{\nu} ) \bigr)^2 \over \sum_{\nu} \sin^2 p_{\nu}}\biggr].\cr}}

This non-locality is of a new type compared with two preceding cases as
$L_{\mu}(p)$ now involves also all $p_{\nu}$ with $\nu \neq \mu$. Now we have
\eqn\gnw{G^W(0) \approx 0.0259, \quad G^W(\pi) \approx - 0.00734, \quad
G^S(0)
\approx 348, \quad G^S(\pi) \approx -159.}
Phase diagrams are shown in Figs. 2($d$), 3($d$), and 4($d$).

$Z_2$: Fig. 2($d$). Domains of the weak and strong coupling regimes are not
overlapped:
\eqn\dswwz{y^W \approx 11.7, \quad y^S \approx 12.6,}
so this case incorporates features of both local and non-local actions.
The phase diagram, except the scale of $y$, has the same features as that for
naive fermion action.

$U(1), \, n_f =2$: Fig. 3($d$). Domain with FI phase appears from $n_f =
1$, expanding with $n_f$. For $n_f = 2$ it appears at
$y \approx 5.19$ and $y \approx 9.72$, so that according to our criterion we
can trust the picture. The phase diagram qualitatively agrees with the result
of Ref. \ref\LMW{L. Lin, I. Montvay, and H. Wittig, Phys. Lett. B264, 407
(1991)
} (there the phase diagrams examined in a region of parameter space that is
different from ours).

$SU(2)$: Fig. 4($d$). The weak and strong coupling domains are determined by
\eqn\dswws{y^W \approx 8.79, \quad y^S \approx 13.2.}
As it happened in all other $SU(2)$ systems, FI phase appears.

To show the relative role of the proper and crossed ladder diagrams, we
display in Figs. 2 (by thin dashed lines) contributions of the proper
ones. This demonstrates that crossed ladder diagrams play the important role
only when the case ($i$) is realized, being in fact negligible in other
cases. We expect that contributions of diagrams that we did not take into
account are actually invisible, at least in the cases ($ii$) and ($iii$).

\newsec{Fermion correlators}

In order to learn more about structure of various phases of Figs. 2 -- 4
we now make a mean field estimation of fermion correlators
$\vev{\psi_m \,\psibar_n}$ for the above systems. We mainly concentrate on an
neighbourhood of the PM-FM critical lines, as it is this domain that is
expected to be the most interesting for the continuum physics.

We shall evaluate the quantity
\eqn\fcor{\vev{\psi_m \,\psibar_n}_{MF} \equiv \big\langle \big[\dsl +  y
\tilde{\Phi}\big]^{-1}_{m n} \big\rangle_{h^*}, } where expectation value in
the r.h.s. is defined in \vevh, while $h^*$ is a solution of the mean field
equation \mfe. Such an expression appears very naturally as a variational
mean
field approximation for the correlator (see
\STs), though there is no strict relation similar to Eq. \MF\ for free
energy.
By definition this is a quenched estimation.

We have the following representations for \fcor:
\eqn\fcorr{\eqalign{\vev{\psi_m & \,\psibar_n}_{MF} \cr
& = \dsl^{-1}_{m \, n} + \sum^{\infty}_{l = 1} (-1)^l  \, y^{l}
\sum_{i_1, \ldots ,i_{l}} \big\langle \dsl^{-1}_{m \, i_1} \,
\tilde{\Phi}_{i_1}
\,
\dsl^{-1}_{i_1 \, i_2} \cdots \tilde{\Phi}_{i_l} \, \dsl^{-1}_{i_l \,
n}\big\rangle_{h^*} \cr
& = {1 \over y} \big\langle \tilde{\Phi}^{\dagger}_{m}\big\rangle_{h^*}
\delta_{m \, n} +
\sum^{\infty}_{l = 1} (-1)^l {1 \over y^{l + 1}}
\sum_{i_1, \ldots ,i_{l-1}} \big\langle \tilde{\Phi}^{\dagger}_{m} \,
\dsl_{m \, i_1} \, \tilde{\Phi}^{\dagger}_{i_1}  \cdots
\dsl_{i_{l-1} \, n} \, \tilde{\Phi}^{\dagger}_{n} \big\rangle_{h^*}. \cr}}
We shall use these two representations for weak and strong coupling regimes
which have been determined by our preceding considerations.

Consider first $\vev{\psi_m \,\psibar_n}_{MF}$ in FM phase in the
approximation
of uncorrelated Higgs fields. Choose their expectation values to be real:
$\vev{\phi^{a}_n}_{h^*} = \delta ^{a\,0} \vev{\phi}$, so that
\eqn\nc{\vev{\tilde{\Phi}_1\,\tilde{\Phi}_2 \cdots \tilde{\Phi}_l}_{h^*} =
\vev{\tilde{\Phi}^{\dagger}_1\,\tilde{\Phi}^{\dagger}_2 \cdots
\tilde{\Phi}^{\dagger}_l}_{h^*} = \vev{\phi}^l.}
Then, from \fcorr\ it follows:
\eqn\fcorws{\eqalign{&\vev{\psi_m \,\psibar_n}^{W}_{MF} = \biggl(\dsl + y
\vev{\phi} \biggr)^{-1}_{m\,n}, \cr
& \vev{\psi_m \,\psibar_n}^{S}_{MF} = \biggl(\dsl + {y \over
\vev{\phi} } \biggr)^{-1}_{m\,n}.}}
This reproduces the well-known result, which has been obtained by various
methods  (for the first references see \ref\JS{J.
Smit, Nucl. Phys. B (Proc. Suppl.) 9, 579 (1989)}, \HN, \Sec), that the
behavior
of the
fermion masses with $\vev{\phi}$ (and, therefore, that of renormalized
$y$) is completely different in the weak and strong coupling regimes. While
in the weak regime one has the usual perturbative Higgs mechanism (with the
Gaussian fixed point for $y$), the fermion masses at strong coupling do not
vanish on the critical lines and in PM phase. In this approximation they tend
to infinity, that is, the fermions decouple there. We shall refer hereafter
to
PM
(FM) phases at weak and strong coupling
regimes as PM(W) and PM(S) [FM(W) and FM(S)], respectively.

In \Zen\ it was contemplated a possibility to use this feature of the
Higgs-Yukawa systems for defining a chiral theory in PM(S) phase of the
mirror
fermion model \mf. The idea was that the mirror fermions $\chi$ decouple
there
leaving behind massless chirally invariant fermions $\psi$. Using the above
approximation, however, it is easy to show that though all goes in such a
way,
the goal cannot be reached: $\psi$ turn to naive fermions.

Although approximation \nc\ yields correct
qualitative picture, it is too rough for a quantitative analysis: it has been
shown by numerical calculations \HN, \HLN, \Boea, that the fermion masses
at strong coupling indeed increase with decreasing $\vev{\phi}$, but remain
finite on the critical line, rather than tend to infinity.

To proceed further we make the mean field estimation of the fermion
condensate
\eqn\fco{\vev{\psibar \, \psi}_{MF} \equiv N^{-D} \sum_n \vev{\psibar_n \,
\psi_n}_{MF}}
along the FM-PM critical lines where it has the form:
\eqn\fcon{\vev{\psibar \, \psi}_{MF} = - 2^{D/2} \, C(y)
\, \vev{\phi} + O(\vev{\phi}^2).}
This allows us to use the ladder approximation, that is, to sum up the
contributions to \fcon\ of the diagrams of Fig. 5 ($a$). Then, from
\cor, \corre\ and \fcorr\ we find

$Z_2:$
\eqn\conz{\eqalign{
&C^{(W)}(y) = {y \, G^W(0) \over 1 + y^2 G^W(0)}, \cr
&C^{(S)}(y) = {y \over y^2 + G^S(0)};}}

$U(1):$
\eqn\conu{\eqalign{
&C^{(W)}(y) =  y\, G^W(0), \cr
&C^{(S)}(y) = {1 \over y}; \cr
}}

$SU(2):$
\eqn\cons{\eqalign{
&C^{(W)}(y) = {2 y \, G^W(0) \over 2 - y^2 G^W(0)}, \cr
&C^{(S)}(y) = {2 y \over 2 y^2 - G^S(0)}. \cr }}
In the case of $U(1)$ only the first diagram of Fig. 5($a$) gives a
contribution to \conu: the situation is very similar to that for critical
lines
\resu\ (see also \STs).

condensates are smooth functions along both PM(W)-FM(W) and PM(S)-FM(S)
critical
lines. In the cases when FI phase appears it prevents us to follow far beyond
the points of intersection of PM-FM and PM-AM lines. The important fact,
however, is that in a neighbourhood of those points condensates remain
smooth functions of both their arguments
$y$ and
$\vev{\phi}$. This is due to the fact that the staggered magnetization
gives no contributions of order $O(\vev{\phi})$ to $\vev{\psibar \,
\psi}_{MF}$ (because of momentum conservation).

In the cases when the PM-FM line is continuous and no FI phase appear, the
function $C(y)$ can be discontinuous at the points of intersection of FM(W)
and
FM(S) lines. We have only three such systems: $Z_2$ with SLAC fermions, and
$U(1)$ with naive and SLAC ($n_f = 1$) fermions [points A in Figs. 2($b$) and
3($a$),($b$), respectively]. Figs. 6 and 7($a$),($b$) show that this
is indeed the case. This means that the condensate, being zero in both PM
phases
and on the whole critical line, is discontinuous in FM phase. The
discontinuity
of the fermion condensate is an evidence of the first order
phase transition. Indeed, the condensate can be defined as the first
order derivative of the free energy in respect to an infinitesimal fermion
mass (see also \ST). As the condensate is an order parameter of the
systems, points A in Figs. 2($b$) and 3($a$),($b$) look like tricritical
points in which the first order phase transition turns to the second order
one.
This allows us to identify the points A with point A in Fig. 2 of Ref. \HN,
thereby considering them as candidates for non-trivial fixed points.

\newsec{Summary and discussion}

We derived the explicit formulae [Eqs. \res\ -- \resul] describing phase
diagrams of a wide class of $Z_2$, $U(1)$, and $SU(2)$ Higgs-Yukawa systems,
and applied them to the systems with naive, SLAC, the Weyl, and mirror
fermion
actions (Figs. 2
-- 4). The phase diagrams turned out to be very different for different
symmetry groups and fermion actions at $\kappa < 0$, being of the same form
at
$\kappa \geq 0$.

The difference between the phase diagrams for different formulation of
lattice
fermions shows a lack of universality in the systems at $\kappa < 0$, and can
be interpreted as a lattice artifact. It is well known that in this region
the
sufficient condition of reflection positivity is not satisfied, and,
therefore, relevance of the systems to well-defined quantum field theories is
under the question. If, however, in that region  physical positivity is
fulfilled, the systems are still interesting from the point of view of
continuum physics. Analysis of scalar propagators at $\kappa < 0$ in
$SU(2)$ system with naive fermions \Bea, \ref\Bocea{W. Bock, A. K. De, C.
Frick, K. Jansen, and T. Trappenberg, Nucl. Phys. B371, 683 (1992)} gives
some
evidence that this is indeed the case.

All the points which are candidates for non-trivial fixed points lies in the
region of negative $\kappa$. There are two types of such points: those
where PM critical lines intersect AM lines forming FI phase, and the points
where we can expect a phase transition separating weak and strong coupling
regimes. The points of the first type present in all $SU(2)$
systems, Figs. 4, and in some of $U(1)$ systems, Figs. 3($c$), ($d$). This
case, however, seems to be excluded: the numerical investigations of such
points in $SU(2)$ system with naive fermions \Boea, \Bea, \Bocea\ gave no
evidence of non-trivial behaviour of the system. Our analysis speaks in
favour
that too: we applied fermionic condensate calculated along the PM-FM
critical lines [Eqs. \fco\ -- \cons] as an order parameter sensitive to first
order phase transitions, and showed that the condensate, as well as both
critical
lines, does not feel these points, remaining smooth functions of their
arguments.

Among the points of the second type only those are relevant to
continuum physics that border on PM phase: at other points either
$\vev{\phi}$
or $\vev{\phi_{st}}$ is not zero. Therefore, the most
interesting are systems which have continuous PM-FM critical lines and no
FI phase. In our examples these are $Z_2$ system with SLAC action and $U(1)$
systems with naive and SLAC ($n_f = 1$) actions. Indeed, in FM phase near
such
points the fermion condensate is discontinuous, Figs. 6 and 7($a$),($b$),
that
is
an  evidence of the first order phase transition separating the FM(W) and
FM(S)
phases. We therefore identify the points A in Figs. 1($b$),
and 2($a$),($b$) with point A in Fig. 2 of Ref. \HN.
If these points are really non-Gaussian, than non-trivial continuum theories
with
two relevant parameters, Higgs and fermion masses, can be defined approaching
the points from FM(W) phase.

The $SU(2)$ systems we considered have no such points. The latter, however,
can exist in $SU(2)$ systems with other formulations of lattice fermions, in
particular, in those with staggered fermions coupled to the Higgs fields in
a local way, when number of the fermions is not too big \STsy, \EK.

We expect that disagreement of the phase diagram for $U(1)$ system with
naive fermion action, Fig. 3($a$), with results of Ref. \HLN, where an
evidence
of FI phase has been found, is due to finite lattice
effects in the numerical calculations. Our results for $SU(2)$ system with
naive fermions are in quantitative agreement with numerical calculations of
Ref. \Boea, and results for $U(1)$ system with mirror fermions are in
qualitative agreement with those of Ref. \LMW\ (we cannot compare them
directly, because the latter have been obtained in different region of
parameter spase of the model). In both cases the appearance of FI phases
has very clear reasons. Therefore we do not see any reasons why our
approximation could fail in this case. The more so, that in this case it is
most stable against $1/D$ corrections (cf. \STs). Otherwise, there must be
some
underlying physics which by unknown reasons is not taken into account by our
approximation, and which worth further investigation. The only change of the
above picture which we cannot exclude is that the FM-PM (perhaps together
with
AM-PM) critical line is actually discontinuous, being teared up by the first
order phase transition line at some value of $y$ not far from $y^A \approx
1.34$. Then
the point $A$ splits into two first order phase transition points where
critical
lines $\kappa_{cr}^{FM(W)}$ and $\kappa_{cr}^{FM(S)}$ end up. In this case,
in
view of the results of Ref. \HJS, one can hardly expect the existence of
non-trivial fixed points in the system. Examination of this issue, however,
requires another technique.

Therefore, it would be very interesting to repeat thorough
numerical investigation of the $U(1)$ system with naive fermion action
near the point ($y^A, \, \kappa^A) \approx (1.34, \, -0.43)$ (for
$n_f = 2$). Another interesting issue is to trace the evolution of the
phase diagrams of such system at finite scalar self-coupling $\lambda$ \HJS\
with increasing $\lambda$.

\bigbreak\bigskip
\centerline{{\bf Acknowledgments}}
\vskip 8pt

We are grateful to H. Yoneyama for enlightening discussions and helpful
suggestions and to S. Sakoda for drawing Figs. 1, 5. S. V. Z. is grateful to
J. L. Alonso and H. Kawai for interesting discussions. It is also a pleasure
for him to thank Elementary Particle Theory Group of Kyushu University, YITP,
and Theory Group of KEK, where parts of this work have been done, for their
kind hospitality. The work of S. V. Z. was partly supported by JSPS.

\listrefs

\bigbreak\bigskip
\centerline{{\bf Figure captions}}
\vskip 8pt

Fig. 1. Diagrams contributed to the expectation value of the fermion
determinant
\d\ to the order $h_n^{2}$; ($a$) is the ladder, ($b$) the crossed ladder
diagrams. Solid lines stand for $\dsl$ or $\dsl^{-1}$, solid circles for
$u'$,
dashed lines for $u''$.

Fig. 2. Phase diagrams of $Z_2$ systems with ($a$) naive, ($b$) SLAC, ($c$)
the Weyl, ($d$) mirror fermion actions. Dashed lines show the contribution of
only diagram of Fig. 1($a$). Point $A$ in ($b$) is discussed in the text
as possible non-trivial fixed  point.

Fig. 3. Phase diagrams of $U(1)$ systems with ($a$) naive ($n_f = 2$), ($b$)
SLAC, ($c$) the Weyl, ($d$) mirror fermion ($n_f = 2$) actions. Dashed lines
in
($d$) is extrapolation of the formulae \resu\ to FI region. Points $A$ in
($a$) and ($b$) are discussed in the text as possible non-trivial fixed
points.

Fig. 4. Phase diagrams of $SU(2)$ systems with ($a$) naive ($n_f = 2$), ($b$)
SLAC, ($c$) the Weyl, ($d$) mirror fermion actions. Dashed lines is
extrapolation of the formulae \resul\ to FI region.

Fig. 5. Diagrams contributed to the fermion condensate \fcon\ to the order
$\vev{\phi}$; ($a$) is the ladder diagrams. Solid circles stand for
$\vev{\phi}$, crosses for site $n$ in \fco; other notations as in Fig. 1.

Fig. 6. Fermion condensate [function $C(y)$ in \fcon] along the PM-FM
critical
line for $Z_2$ system with SLAC fermions.

Fig. 7. The same as in Fig. 6, but for $U(1)$ systems with ($a$) naive, ($b$)
SLAC
fermions.

%%% insret figs. %%%
\vfil\eject

\font\cour=cmtt12 scaled\magstep2

\epsfysize=0.4\hsize
\centerline{\epsfbox{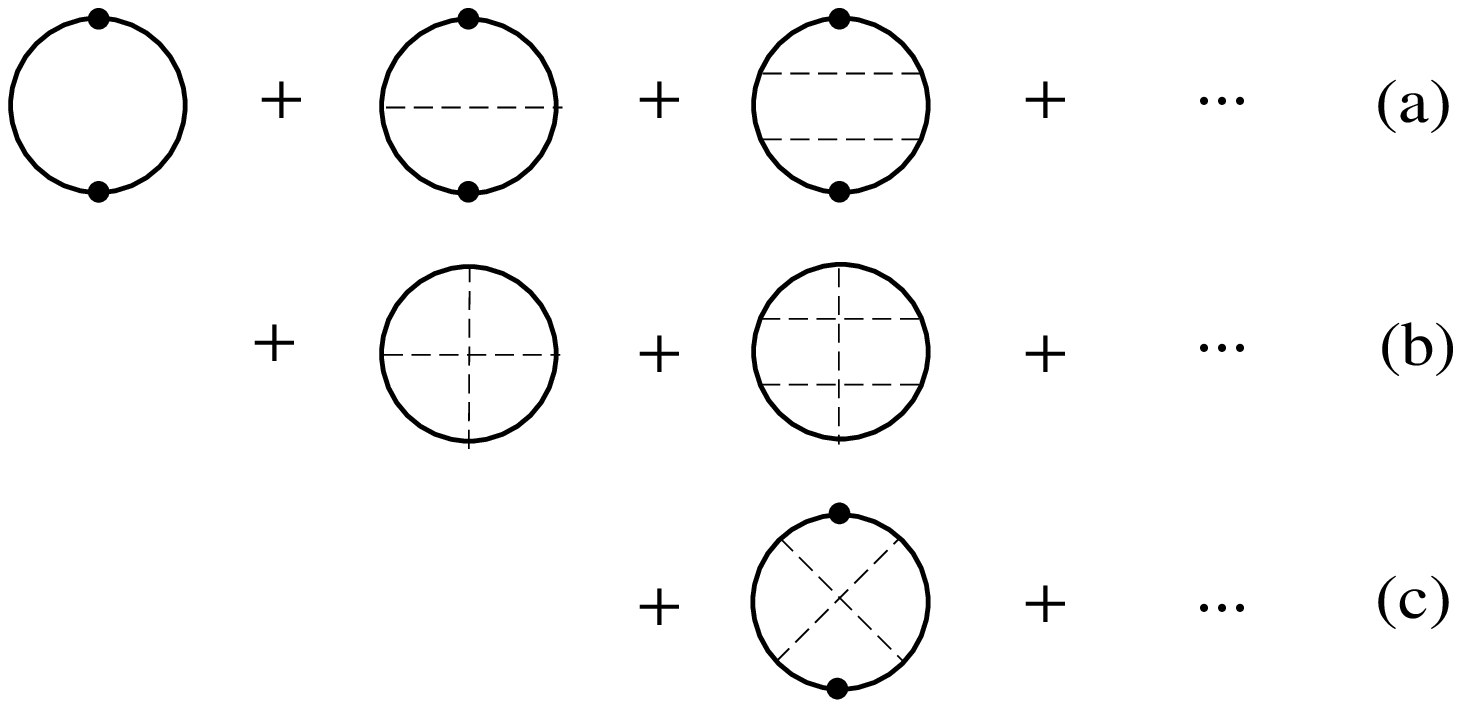}}
\vskip2em
\centerline{{\cour Fig. 1}}
\vfil\eject

\epsfysize=0.6\hsize
\centerline{\epsfbox{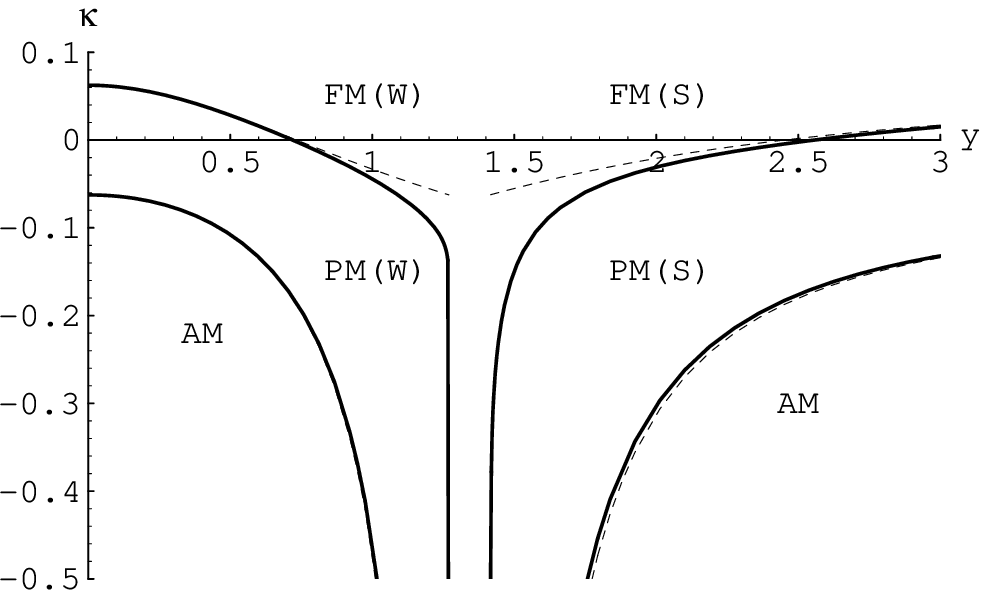}}
\centerline{{\cour Fig. 2(a)}}
\vskip2em

\epsfysize=0.6\hsize
\centerline{\epsfbox{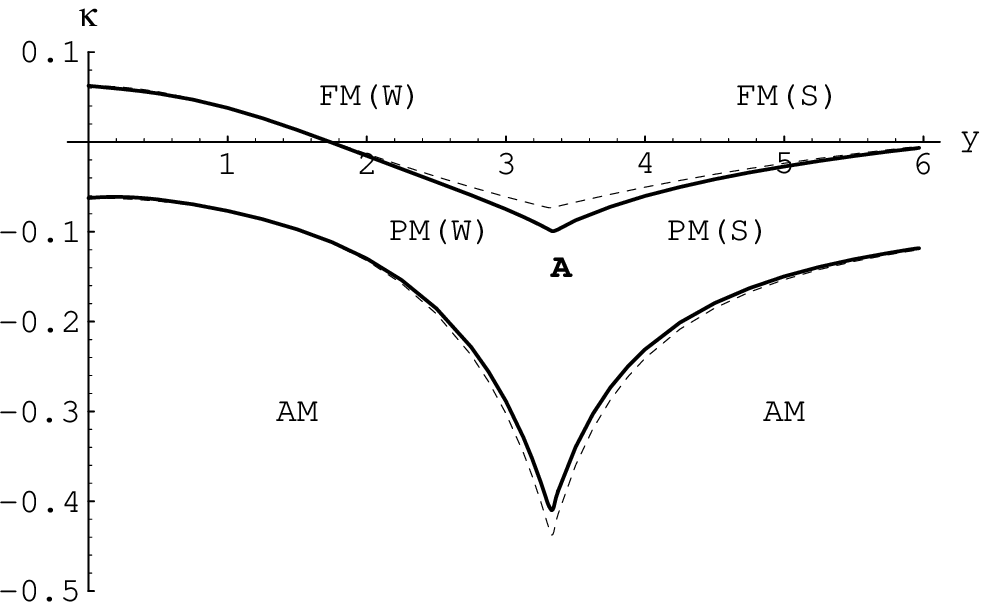}}
\centerline{{\cour Fig. 2(b)}}
\vfil\eject

\epsfysize=0.6\hsize
\centerline{\epsfbox{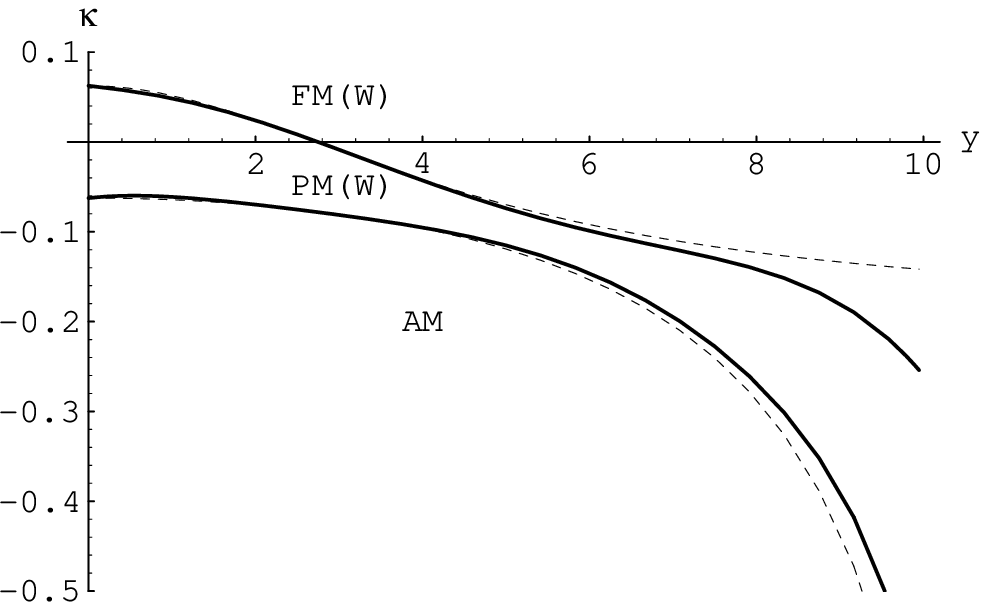}}
\centerline{{\cour Fig. 2(c)}}
\vskip2em

\epsfysize=0.6\hsize
\centerline{\epsfbox{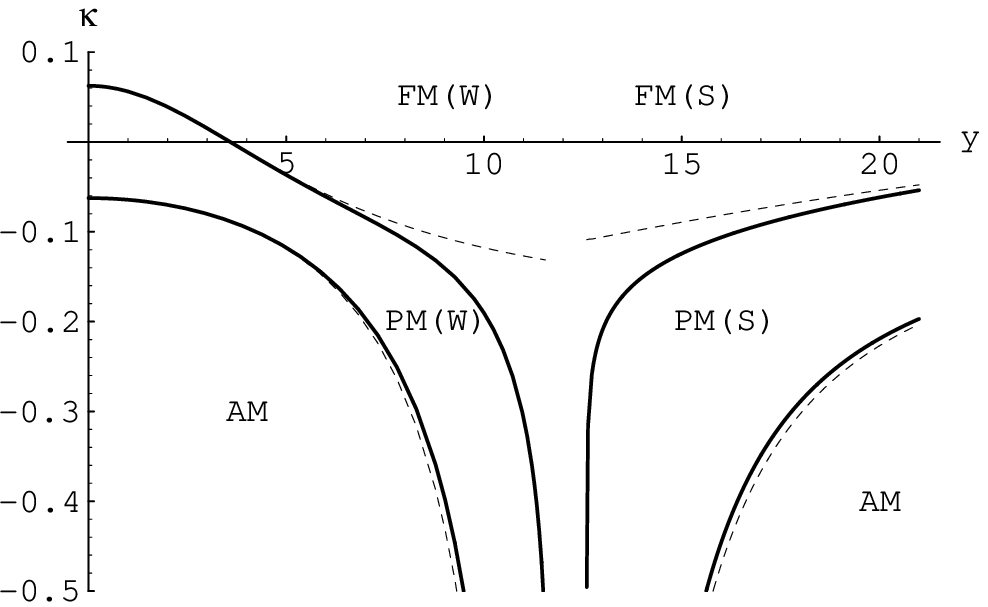}}
\centerline{{\cour Fig. 2(d)}}
\vfil\eject

\epsfysize=0.6\hsize
\centerline{\epsfbox{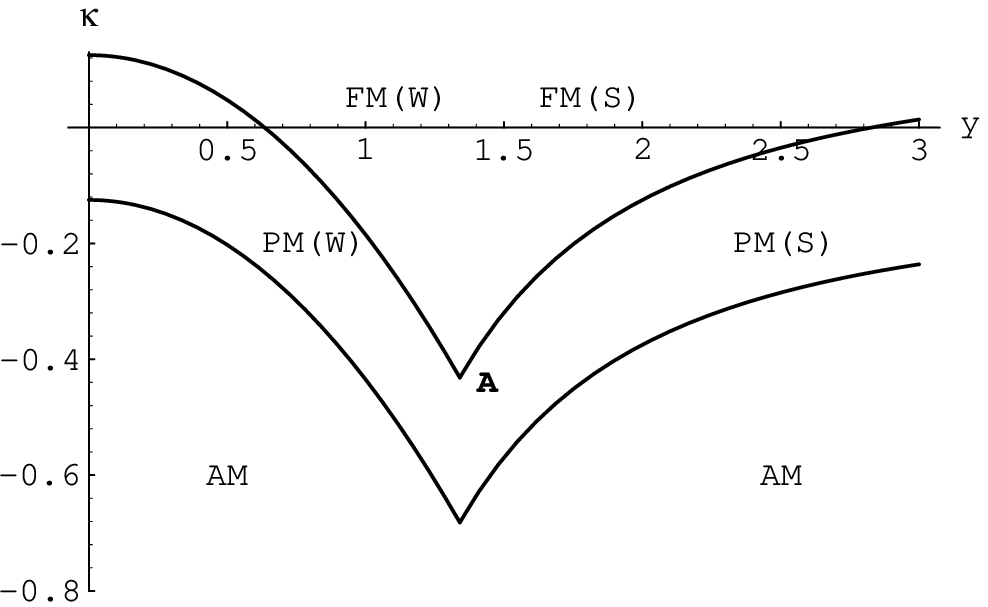}}
\centerline{{\cour Fig. 3(a)}}
\vskip2em

\epsfysize=0.6\hsize
\centerline{\epsfbox{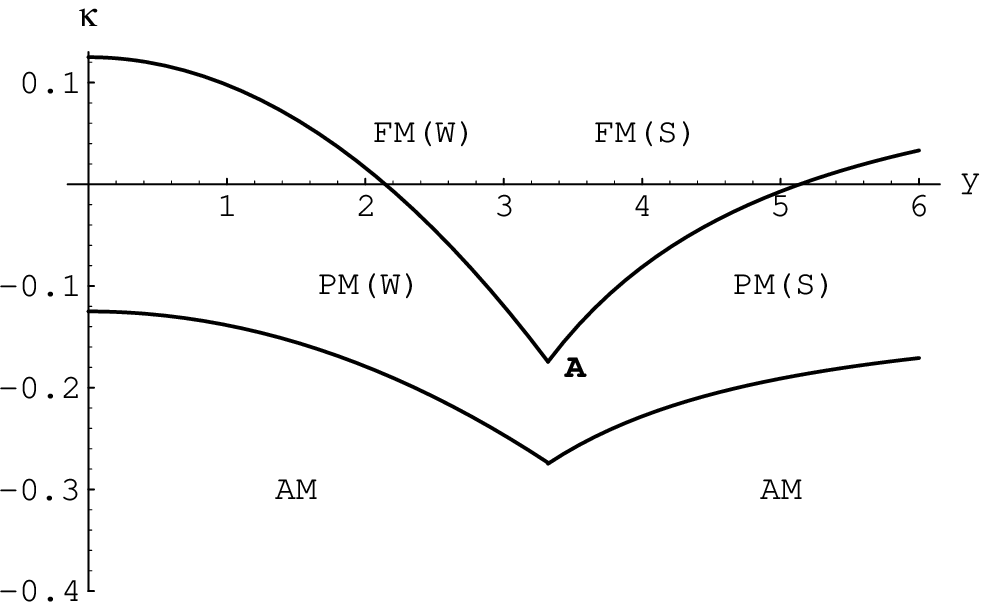}}
\centerline{{\cour Fig. 3(b)}}
\vfil\eject

\epsfysize=0.6\hsize
\centerline{\epsfbox{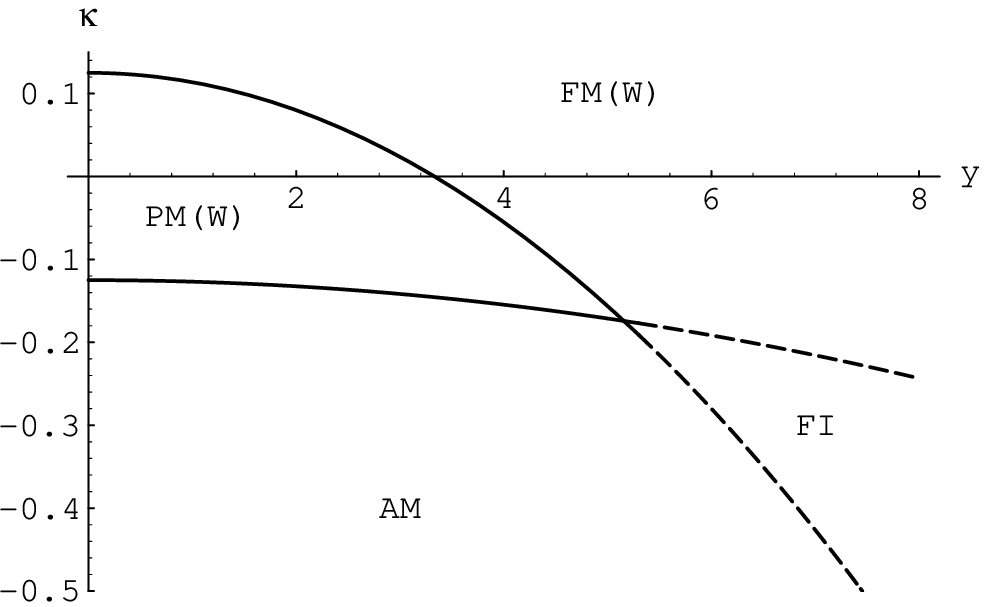}}
\centerline{{\cour Fig. 3(c)}}
\vskip2em

\epsfysize=0.6\hsize
\centerline{\epsfbox{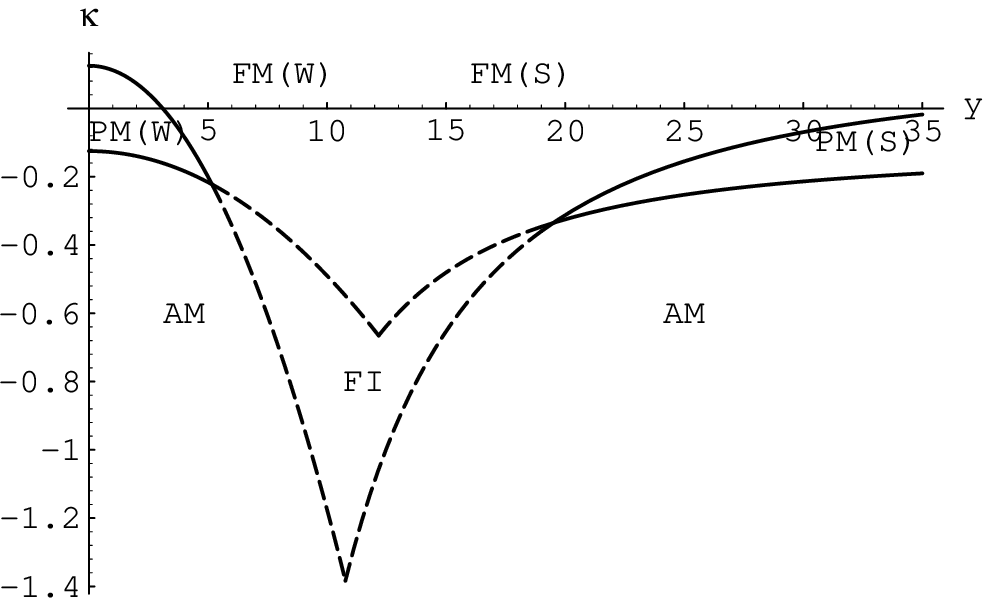}}
\centerline{{\cour Fig. 3(d)}}
\vfil\eject

\epsfysize=0.6\hsize
\centerline{\epsfbox{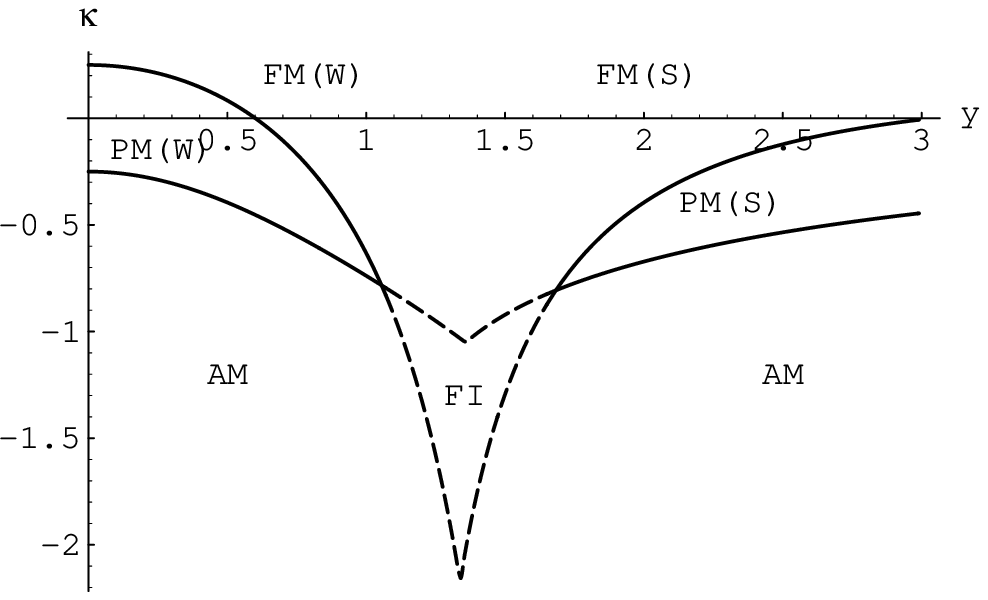}}
\centerline{{\cour Fig. 4(a)}}
\vskip2em

\epsfysize=0.6\hsize
\centerline{\epsfbox{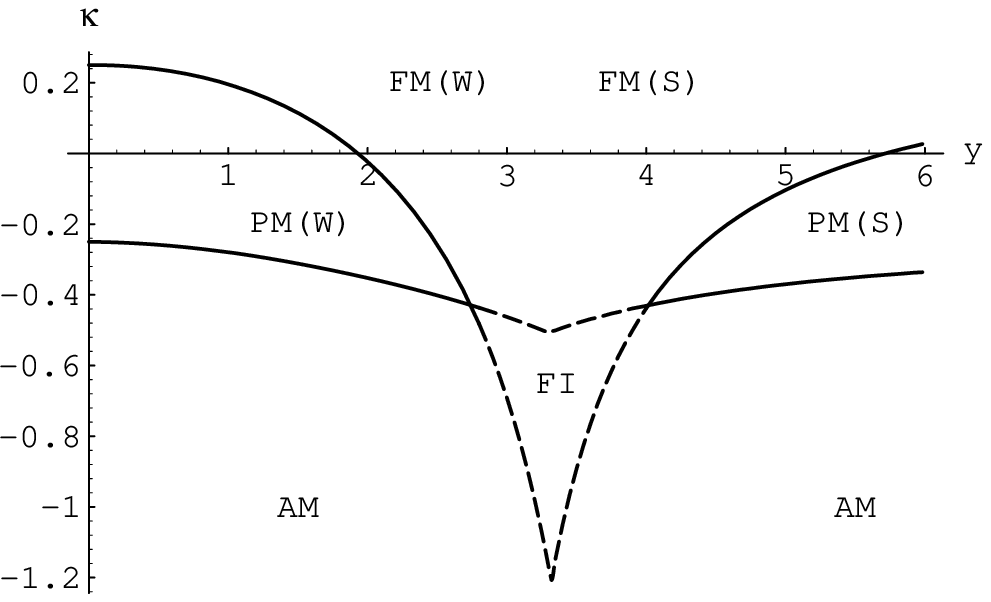}}
\centerline{{\cour Fig. 4(b)}}
\vfil\eject

\epsfysize=0.6\hsize
\centerline{\epsfbox{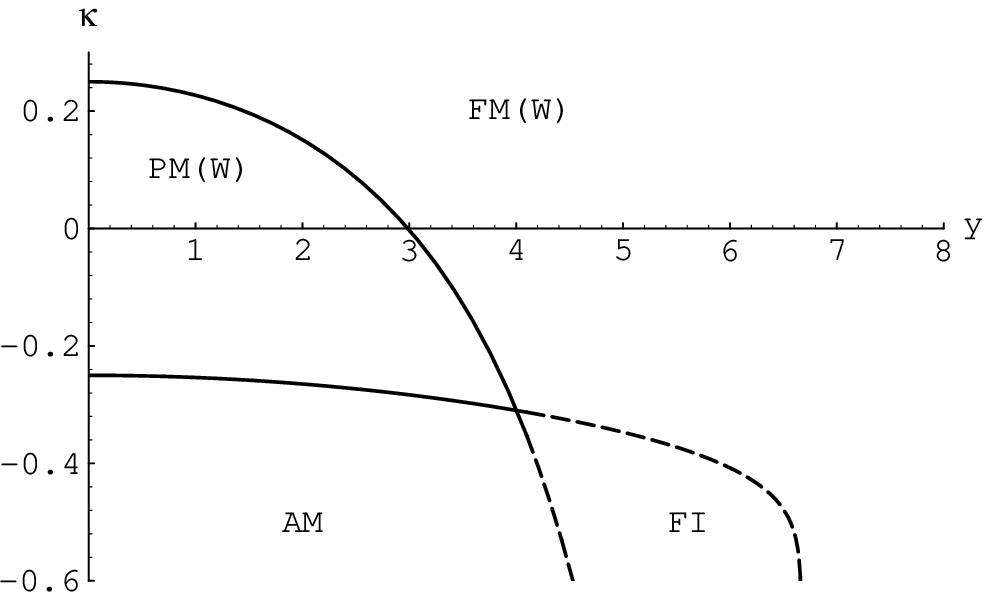}}
\centerline{{\cour Fig. 4(c)}}
\vskip2em

\epsfysize=0.6\hsize
\centerline{\epsfbox{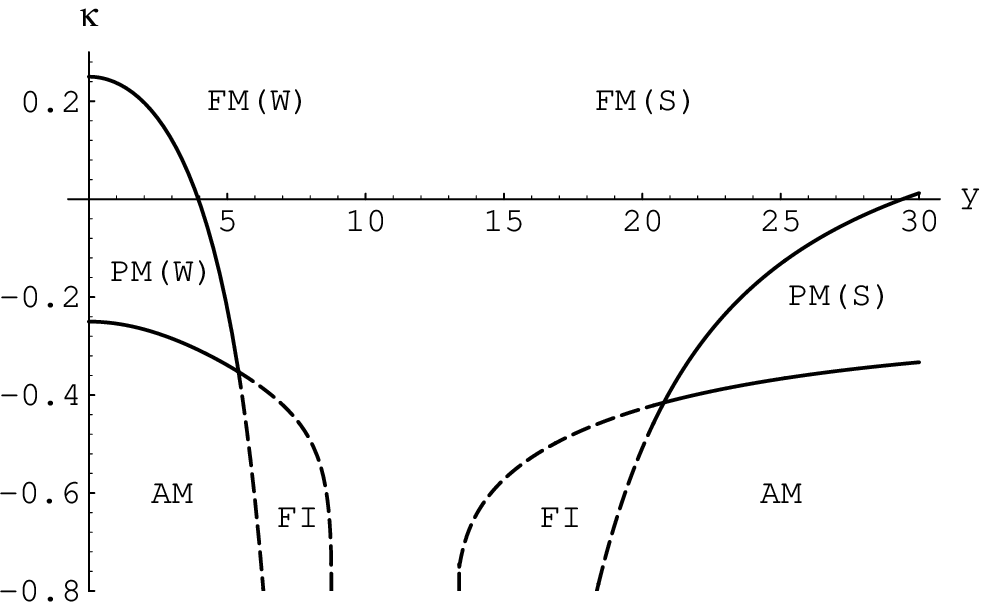}}
\centerline{{\cour Fig. 4(d)}}
\vfil\eject

%\vskip4em
\epsfysize=0.3\hsize
\centerline{\epsfbox{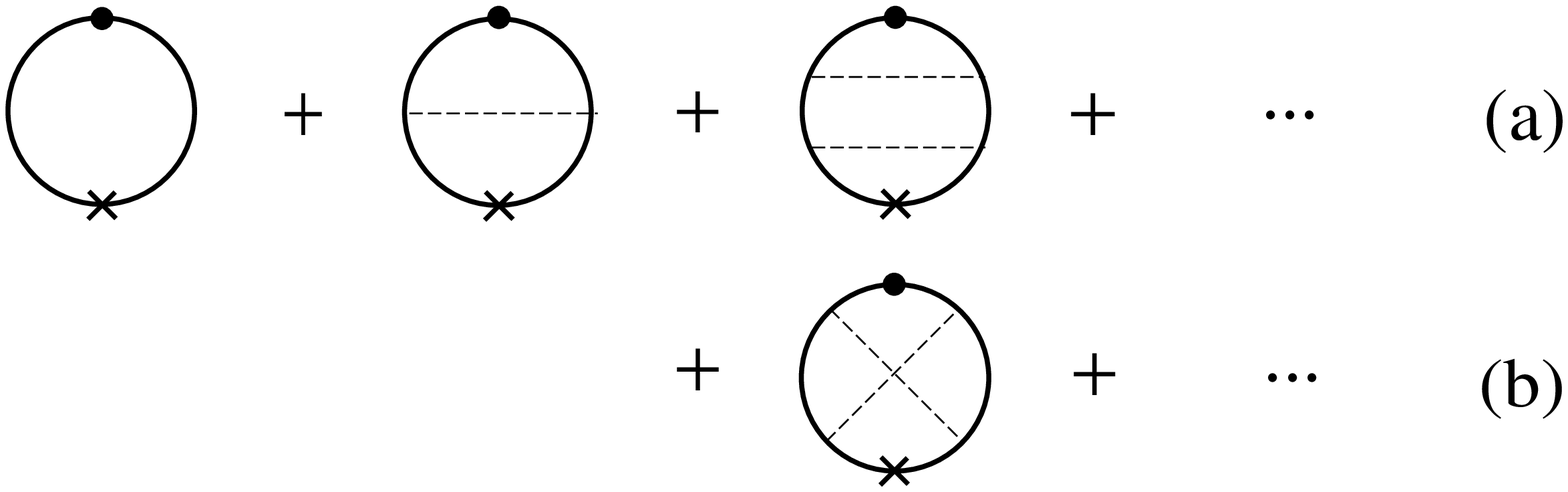}}
\vskip2em
\centerline{{\cour Fig. 5}}
\vskip10em

\epsfysize=0.55\hsize
\centerline{\epsfbox{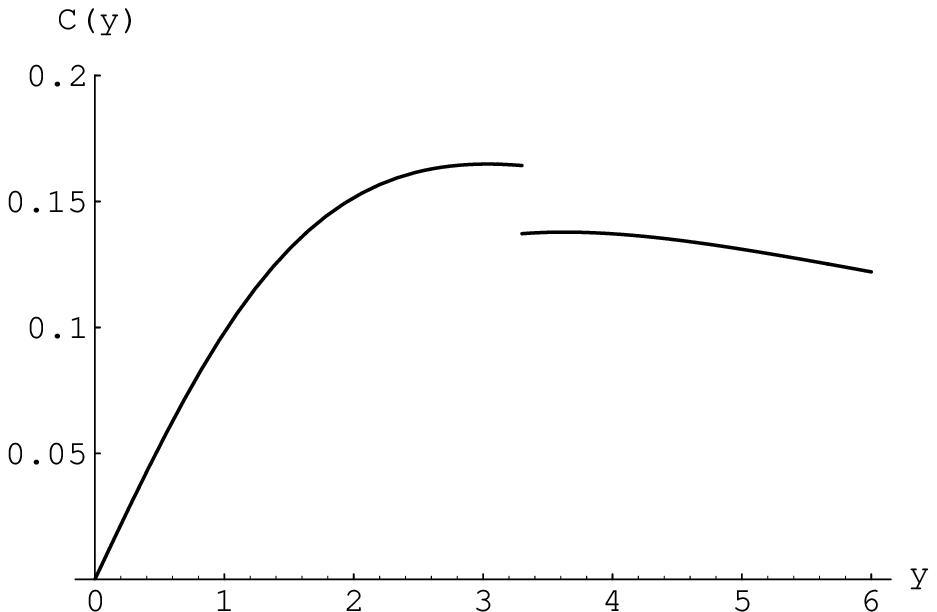}}
\vskip1em
\centerline{{\cour Fig. 6}}
\vfil\eject

\epsfysize=0.55\hsize
\centerline{\epsfbox{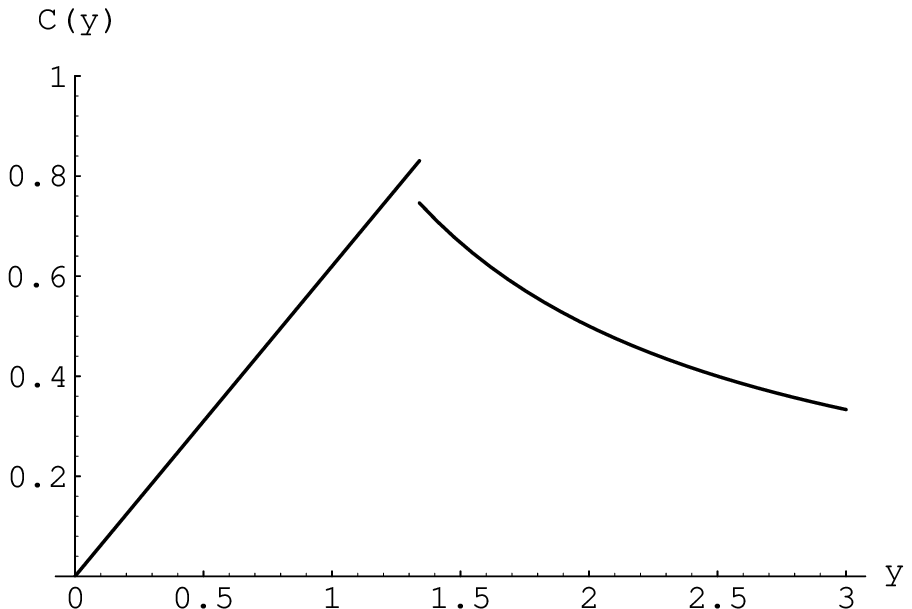}}
\vskip1em
\centerline{{\cour Fig. 7(a)}}
\vskip2em

\epsfysize=0.55\hsize
\centerline{\epsfbox{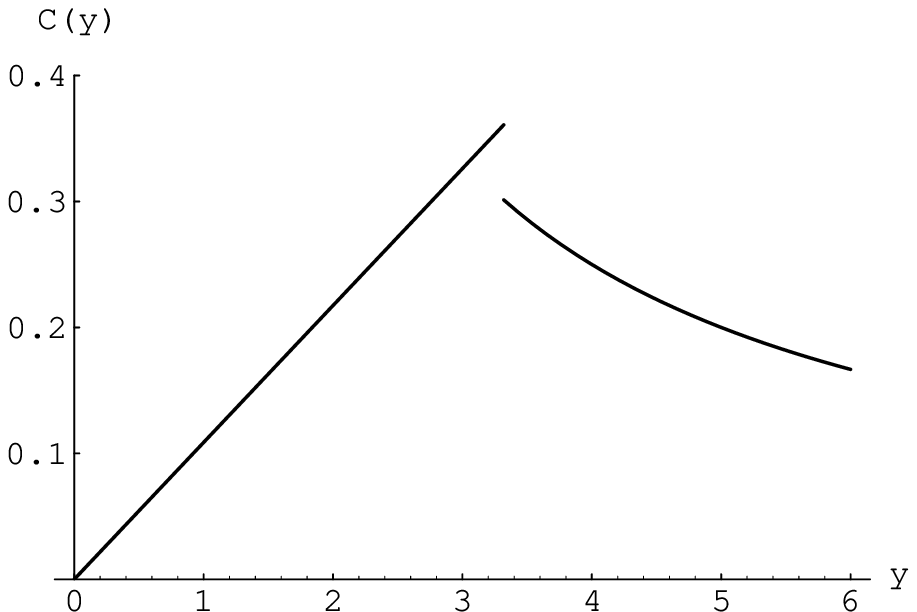}}
\vskip1em
\centerline{{\cour Fig. 7(b)}}
\vfil\eject
%%% end %%%

\bye